\documentclass[12pt]{article}

\usepackage{graphicx} 
\usepackage[affil-it]{authblk}
\usepackage[]{tikz}
\usetikzlibrary{arrows}

\begin{document}

\title{miR-34a-5p and miR-34a-3p contribute to the signaling pathway of p53 by targeting overlapping sets of genes}
\author{Abderrahim Chafik}
\affil{ Minist$\grave{e}$re de l'Emploi et des Affaires Sociales, Rue Al Jommayz, Hay Riad , Rabat, Morocco\\
The Abdus Salam International Centre for Theoretical Physics (ICTP), Strada Costiera, 11
		I - 34151 Trieste Italy}

\maketitle

\abstract{In contrary to the common belief that only one strand of the pre-miRNA is active (usually the 5p one that is the more abundant) while the second one (miRNA*) is discarded, functional 5p and 3p have been observed for many miRNAs. Among those miRNAs is miR-34a which is a target gene of the tumor suppressor p53. In this paper we have re-examined the role of miR-34a-5p and miR-34a-3p in the signaling pathway of p53. We found that they target overlapping sets of genes (MDM2 and THBS1). By a GO enrichment analysis  we found that THBS1 is involved in cancer and metastasis relevant processes. We have also deduce that p53, MDM2 and miR-34a are linked by a type 1 incoherent FFL that represent a novel mechanism for accelerating the response of p53 to external stress signals.}

\section{Introduction}

p53 has been dubbed "guardian of the genome" \cite{key1} or gatekeeper \cite{key2}  due to its central role in  maintaining the genomic stability and tumor suppression \cite{key3, key4, key5}. It has been found to be mutated in about half of the human cancers \cite{key6, key7}. Since its discovery p53 has been subject to a tremendous amount of work making it one of the most extensively studied gene. Its tumor suppressive role consists in inducing anti-proliferative cellular responses to a variety of stress signals, namely a cell-cycle arrest, senescence or an apoptosis. p53 can then be activated in response to DNA damage,  hypoxia, or aberrant growth signals resulting from deregulated expression of oncogenes \cite{key8, key9, key10}. Under normal conditions p53 is expressed at low level as a result of the action of the ubiquitin-protein ligase MDM2, that repress p53 through ubiquitin-mediated degradation \cite{key11, key12}. In response to a stress signal as DNA damage p53 undergoes post-transcriptional modifications (phosphorylation and acetylation) by various kinases, including ATM, ATR, DNA-PK, Chk1, and Chk2 \cite{key13}. These modifications activate p53 by inhibiting MDM2 from binding to its N-terminal and allow it to carry out its major function as a transcription factor that binds to specific DNA sequences and activate the transcription of adjacent genes. By inducing the expression of its target genes p53 induces cell-cycle arrest or apoptosis.

MicroRNAs (miRNAs) are small (20 – 25 nucleotides) non-coding RNAs that negatively regulate gene expression post transcriptionally by base-pairing to the 3'UTR of target mRNAs leading to repression of protein production or mRNA degradation (For a recent introduction see for example \cite{key14}). a single miRNA can target hundreds of different mRNAs, and a single mRNA can be coordinately suppressed by multiple different miRNAs, They are involved in many biological processes, in particular cancer-relevant processes such as proliferation, cell cycle control, apoptosis, differentiation, migration and metabolism. A miRNA can have either oncogenic or tumor suppressive function. Examples of tumor suppressor miRNAs are let-7 family, miR-29, miR-34 and miR -15, and of oncogene miRNAs are miR-17-92, miR-155 and miR-221.

The biogenesis of miRNAs can be summarized as follows: a primary transcript (pri-miRNA) is first generated by RNA polymerase II as separate transcriptional units or embedded within the introns of protein coding genes. Then the  primary transcript is processed by  the microprocessor complex containing the RNase III enzyme Drosha to an approximate 70-nucleotide (nt) pre-miRNA hairpin (the precursor-miRNA) \cite{key15, key16, key17}. Pre-miRNAs are subsequently exported to the cytoplasm by exportin 5 (XPO5) \cite{key18, key19} where their terminal loops are excised by the RNase III Dicer to give rise to a double-stranded ~22 nt stem composed of 5'′ and 3'′ strands representing 5p and 3p respectively. While one of the two strands (the passenger strand miRNA*) is discarded, the other one (the guide) miRNA is then embeded into the RISC (RNA Induced Silencing Complex) to complementary target mRNA for post-transcriptional gene silencing \cite{key20}.

Since it was believed that according to the thermodynamic stability of the pre-miRNA cells preferentially select the less stable one of the two strands (the guide) and destroy the other one (the miRNA*), early works miRNAs have focused on the guide strand (which was usually considered as the 5p one because it was found to be more abundant than its counterpart miRNA* in humans \cite{key21}). However, even though miRNA* are less abundant they are often present and remain functional because they conserve their seed sequences  and have been isolated from RISC \cite{key22, key23} . It has been shown by profiling analyses that both strands could be co-accumulated in some tissues while being subjected to strand selection in others \cite{key24, key25, key26}. The interplay between the 5p and 3p strands from the same precursor has been shown either in arm-switching where the dominant miRNA is switched from one arm of the precursor to the other \cite{key27} or by targeting overlapping sets of genes when the abundances of the two strands are similar \cite{key28}. Functional 5p and 3p have been characterized for many  miRNAs. Examples include miR-9 \cite{key29}, miR-17 \cite{key30}, miR-19 \cite{key23}, miR-28 \cite{key31}, miR-30c \cite{key32}, miR-125a \cite{key33}, miR-142 \cite{key34}, miR-155 \cite{key35}, miR-199 \cite{key36, key37}, miR-223 \cite{key38}, miR-342 \cite{key39}, miR-2015 \cite{key40}, miR-18a \cite{key41}, and miR-582 \cite{key42}. In some cases the two arms function in opposing ways (e.g. miR-28 \cite{key31} and miR-125 \cite{key33}), while in others they function in joint fashion (e.g. miR-199 \cite{key37} and miR-155 \cite{key35}).

miRNAs have been shown to be important components in the p53 network  Their interactions with p53 have been demonstrated through the identification of several miRNAs as direct target genes of p53. By inducing the expression of specific miRNAs that have a tumor suppressive function a novel mechanism for tumor suppression for p53 has been then revealed. In particular the role of the miR-34 family has been reported in several studies \cite{key43}.

In mammalians miR-34 family consists of miR-34a, miR-34b and miR-34c, that are encoded by two different genes. miR-34a is encoded by an individual transcript in chromosome 1 and expressed in a majority of tissues, while miR-34b and miR-34c share a common primary transcript in chromosome 11 and are mainly expressed in lung tissues. Several studies have reported that the members of miR-34 family were direct target genes of p53 and their upregulation induces apoptosis and cell-cycle arrest.  \cite{key44, key45, key46, key47, key48, key49, key50}. Indeed, Ectopic expression of miR-34 induces cell cycle arrest in both primary and tumour-derived cell lines \cite{key50}. Inactivation of miR-34a strongly attenuates p53-mediated apoptosis in cells exposed to genotoxic stress \cite{key46}. mir-34b/mir-34c were also down-regulated in p53-null human ovarian carcinoma cells and both cooperate in suppressing proliferation of neoplastic epithelial ovarian cells \cite{key45}. Since cell-cycle arrest and apoptosis are the responses of p53 to the stress signals, these facts imply that miR-34 mediate the tumor suppressive function of p53.  On the other hand the members of the miR-34 family can have decreased expression in cancer because of the inactivating mutations of p53 or the expression of viral inhibitors of p53, but also as a consequence of their own mutational or epigenetic inactivation.

Since 30\% of all genes and the majrity of the genetic pathways are regulated by miRNAs \cite{key51, key52, key53}, we can expect some miRNAs to regulate p53 and its pathway. This hypothesis has been verified and some miRNAs have been identified as regulators of p53. miR-504 can negatively regulate p53 expression through its binding to two binding sites in human p53 3′-UTR \cite{key54}. It has also been reported that miR-125b is another miRNA targeting p53 \cite{key55}. Another regulator is miR-29 that was identified as a positive regulator of p53 that up-regulate p53 protein levels and induce p53-mediated apoptosis through repression of p85α \cite{key56}. Furthermore, miR-34a, which is a transcription target of the p53 protein, was also found to positively regulate p53 activity and function in apoptosis through its direct negative regulation of SIRT1 \cite{key57}. SIRT1 is a negative regulator of p53, which physically interacts with p53 and deacetylates Lys382 of p53 \cite{key58}.

The purpose of this paper is the study of the role of both miR-34a-5p and miR-34a-3p in the signaling pathway of the p53.

\section{Materiels and methods}

We begin by using DIANA-miRPath v3. 0 which is a miRNA pathway analysis web-server, providing accurate statistics utilizing predicted or experimentally validated miRNA interactions derived from DIANA-TarBase \cite{key59}. We first perform a KEGG reverse search for miRNAs in the p53 signaling pathway (|hsa04115) by choosing the TarBase v 7.0 for method. In the same web-server we obtain the target genes of the different miRNAs that we found. The distribution of the target genes in the pathway can also be displayed. 

Then we use the cytoscape 3.4.0 software to visualize the interactions where the relevant genes are involved. We take the intAct \footnote{IntAct is one of the largest available repositories for curated molecular interactions data, storing PPIs as well as interactions involving other molecules. It is hosted by the European Bioinformatics Institute. IntAct has evolved into a multisource
	curation platform and many other databases curate into IntAct and make their data available through it \cite{key23}. } . to be our source of interactions. We filter by taxonomy identifier to restrict the obtained network to the human case. In order to form clusters of genes that share a similar function we use the  clusterMaker a Cytoscape application with the GLay Community Clustering algorithm. the BiNGO a Cytoscape Plugin \cite{key60} gives the Gene Ontology (GO) terms that are significantly overrepresented in each cluster.

\section{Results and discussion}

\begin{figure}
	\centering
	\includegraphics[width=0.8\textwidth]{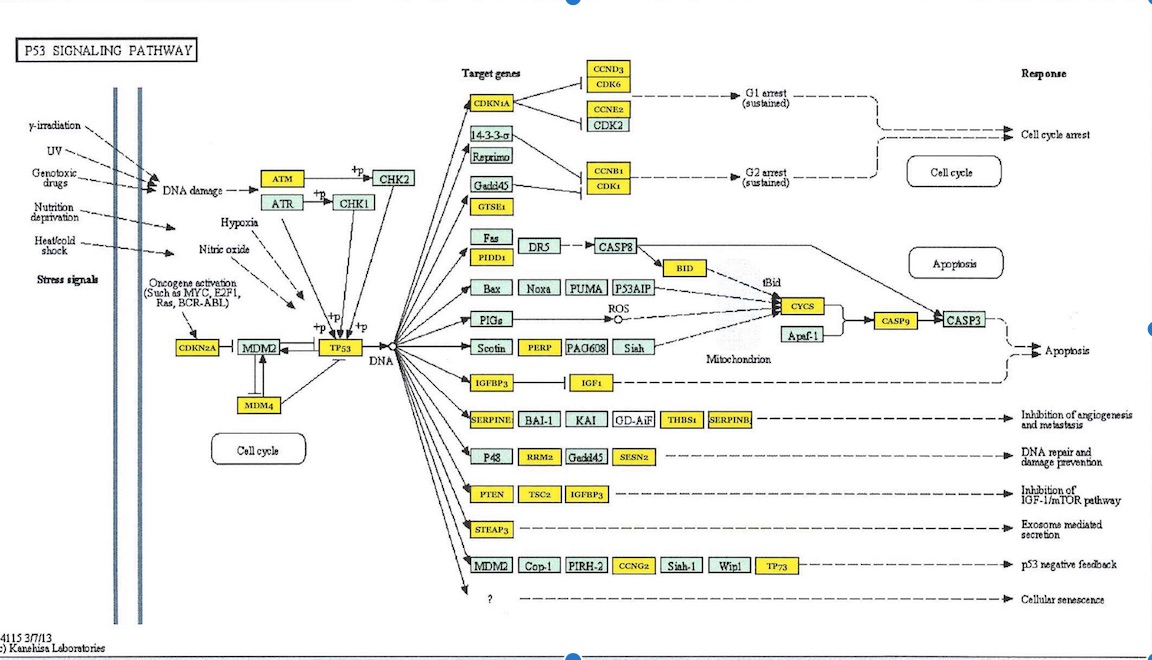}
	\caption{The signaling pathway of p53 with the target genes of miR-34-5p/3p highlighted}
	\label{fig:path}
\end{figure}

The result of the KEGG reverse search was a list containing 660 different miRNAs including  hsa-miR-34a-5p (p-value =  1.840530e-100) and  hsa-miR-34a-3p (p-value =  4.325757e-5). The respective p-values can be interpreted as expressing the fact that 5p is more abundant than 3p. The list of target genes contains 30 genes for miR-34a-5p and two for miR-34a-3p (MDM2 and THBS1) that are also target genes of the miR-34a-5p. The involvement of these genes in the p53 signaling pathway is shown in figure \ref{fig:path}.

\subsection*{THBS1}

\begin{figure}
	\centering
	\includegraphics[width=\textwidth]{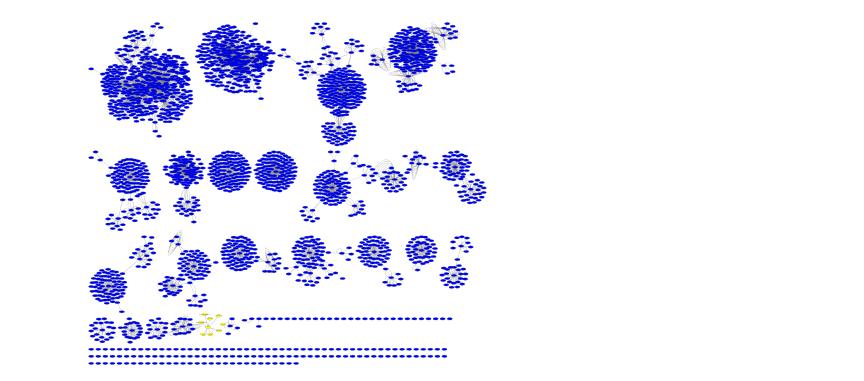}
	\caption{The result of the clustering of the initial network representing different interactions of the target genes of miR-34a-5p/3p}
	\label{fig:cluster}
\end{figure}

\begin{figure}
	\centering
	\includegraphics[width=\textwidth]{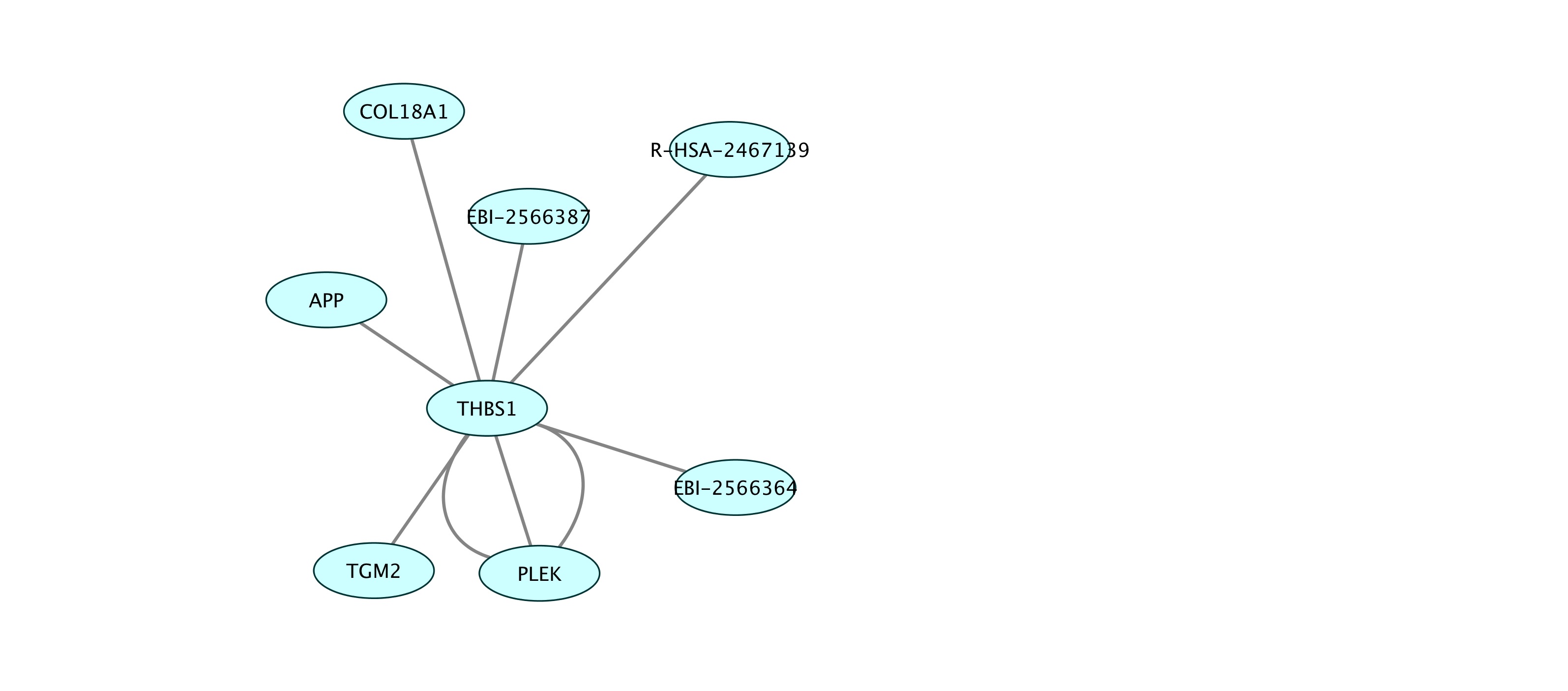}
	\caption{The cluster representing THBS1 and its interactions}
	\label{fig:thbs1}
\end{figure}

By loading the list of the 30 target genes of miR-34a-5p (containing also the two targets of miR-34a-3p) in Cytoscape 3.4.0 and searching the interaction in IntAct we obtain a network composed of 2492 nodes and 6493 edges. After clustering this network by the clusterMaker application we obtain the result shown in figure \ref{fig:cluster} where the cluster containing THBS1 was highlighted, This cluster is shown in figure \ref{fig:thbs1}.

In order to find the pathways in which the THBS1 gene is involved we use the BINGO application of Cytoscape, and the relevant GO terms that are overexpressed in the corresponding cluster are given in table \ref{table:GO}

\begin{table}
	\centering
	\begin{tabular}{l c}
			\hline
     	\textbf{GO terms overexpressed} & \textbf{p-value} \\
	      	\hline
		Regulation of response to external stimulus & $2.34 \times 10^{-6}$ \\
			\hline
		Regulation of response to stress					&  $3.09\times10^{-6}$   \\
			\hline
		Regulation of cell adhesion							 &  $6.29\times10^{-4}$     \\
			\hline
		Positive regulation of apoptosis			   &  $5.48\times10^{-3}$     \\
			\hline
		Negative regulation of cell-matrix adhesion  &  $3.07\times10^{-3}$   \\
			\hline
		Induction of apoptosis									 &  $1.42\times10^{-3}$     \\
			\hline
		Negative regulation of angiogenesis				&   $6.48\times10^{-3}$    \\
		\hline
	\end{tabular}
	\caption{GO term enrichment for the THBS1}
	\label{table:GO}
\end{table}

According to this analysis we can deduce that the THBS1 gene is involved in regulating tumor suppression processes. But it is also clear that it contributes to some anti metastatic pathways like angiogenesis and cell adhesion.

\subsection*{MDM2}

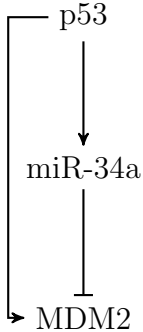
\begin{figure}
	\begin{tikzpicture}[ ->,>=stealth']
	
	\path (2,5) 	node	(p53) {p53};
	\path (2,3) 	node	(miR-34a) {miR-34a};
	\path (2,1) 	node	(MDM2) {MDM2};
	
	\draw[thick, ->]	(p53) -- (miR-34a);
	\draw[thick, -|]	(miR-34a) -- (MDM2);
	\draw[thick]		  (p53) -- (1,5) -- (1,1) -- (MDM2);
	
	\end{tikzpicture}
	\caption{Type 1 incoherent FFL composed of p53, miR-34a and MDM2 }
	\label{fig:ffl}
\end{figure}

As was mentioned in the introduction MDM2 and p53 are linked by a negative feedback loop. However, since MDM2 is a target gene of miR-34a which means that it is negatively regulated by this miRNA, we can represent the interaction between the three molecules (p53, MDM2 and miR-34a) as a Feed Forward Loop (FFL). According to the signs of the three interactions (activation/repression) the present loop is a type 1 incoherent FFL (figure \ref{fig:ffl}).

The dynamics of this FFL can be described as follows: when p53 begins to accumulate as a response to stress signals MDM2 first rises since it is positively regulated by this gene. miR-34a is one of the direct target genes of p53, thus, at some level of p53 the expression of miR-34a begins also to increase, when it reaches some cellular level MDM2 begins to decrease. It has been demonstrated that the response time of the type 1 incoherent FFL is smaller than the one of a single regulation system \cite{key61}. 

Thus in cases where speedy responses are needed this type of regulatory loop is more advantageous than the simple one. In our case MDM2 is the principal cell antagonist of p53 that limits its anti growth function in normal cells. Indeed, in unstressed cells p53 is present at very low levels due to the continuous degradation by ubiquitination which is mediated by MDM2. In stressed cells, in order for p53 to accumulate in response to stress signals it has to escape the degradation by MDM2. This is achieved by some protein kinases that phosphorylate its amino-terminal sites which are close to the MDM2 binding regions of the protein, inhibiting the interaction between the two proteins. on the other hand MDM2 itself is the product of p53-inducible gene. Thus the two genes are linked by an autoregulatory negative feedback loop  (\cite{key62}and references therein).
Thus the FFL constitute an other mechanism that accelerates the stabilization of p53 in response to external stimilus.

\section*{Acknowledgment}
I wish to thank the Abdus Salam ictp for hospitality.

\bibliographystyle{unsrt}
\bibliography{chafikbib}

\end{document}